\documentclass[conference]{IEEEtran}
\IEEEoverridecommandlockouts
\usepackage{cite}
\usepackage{amsmath,amssymb,amsfonts}
\usepackage{algorithmic}
\usepackage{graphicx}
\usepackage{textcomp}
\usepackage{subcaption}
\usepackage{xcolor}
\def\BibTeX{{\rm B\kern-.05em{\sc i\kern-.025em b}\kern-.08em
    T\kern-.1667em\lower.7ex\hbox{E}\kern-.125emX}}

\usepackage{dsfont}
\usepackage{tikz}
\usetikzlibrary{shapes, positioning, matrix, calc, arrows}
\definecolor{tikzblau}{RGB}{93,133,195}
\usepackage[nolist]{acronym}
\usepackage{pgfplots}
\pgfplotsset{compat=1.15}

\newcommand{\lath}[1]{\ensuremath{\lambda_{\text{#1}}}}

\makeatletter
 \let\old@ps@headings\ps@headings
 \let\old@ps@IEEEtitlepagestyle\ps@IEEEtitlepagestyle
 \def\confheader#1{%
 \def\ps@headings{%
 \old@ps@headings%
 \def\@oddhead{\strut\hfill#1\hfill\strut}%
 \def\@evenhead{\strut\hfill#1\hfill\strut}%
 }%
 \def\ps@IEEEtitlepagestyle{%
 \old@ps@IEEEtitlepagestyle%
 \def\@oddhead{\strut\hfill#1\hfill\strut}%
 \def\@evenhead{\strut\hfill#1\hfill\strut}%
 }%
 \ps@headings%
 }
 \makeatother

\confheader{%
PUBLISHED In 2024 International Conference on Electrical Machines (ICEM), 1–7. 2024, doi:10.1109/ICEM60801.2024.10700088
 }

\begin{document}

\title{Thermal Model Calibration of a Squirrel-Cage Induction Machine \\
\thanks{This work is funded by the Deutsche Forschungsgemeinschaft (DFG, German Research Foundation) – Project-ID 492661287 – TRR 361 and the Research Training Group 2128, the Athene Young Investigator Program of TU Darmstadt, the Graduate School Computational Engineering at TU Darmstadt.}
}

\author{\IEEEauthorblockN{1\textsuperscript{st} Leon Blumrich}
\IEEEauthorblockA{\textit{Institute for Electrical Drive Systems (EAS)} \\
\textit{Technische Universität Darmstadt}\\
Darmstadt, Germany \\
leon.blumrich@tu-darmstadt.de}
\and
\IEEEauthorblockN{2\textsuperscript{nd} Christian Bergfried}
\IEEEauthorblockA{\textit{Institute for Accelerator Science}\\
\textit{ and Electromagnetic Fields (TEMF)} \\
\textit{Graduate School CE}\\
\textit{Technische Universität Darmstadt}\\
Darmstadt, Germany \\
christian.bergfried@tu-darmstadt.de}
\and

\IEEEauthorblockN{3\textsuperscript{rd} Armin Galetzka}
\IEEEauthorblockA{\textit{Institute for Accelerator Science}\\\textit{ and Electromagnetic Fields (TEMF)}  \\
\textit{Technische Universität Darmstadt}\\
Darmstadt, Germany}
\and
\IEEEauthorblockN{4\textsuperscript{th} Herbert De Gersem}
\IEEEauthorblockA{\textit{Institute for Accelerator Science}\\
\textit{ and Electromagnetic Fields (TEMF)} \\
\textit{Graduate School CE}\\
\textit{Technische Universität Darmstadt}\\
Darmstadt, Germany}
\and
\IEEEauthorblockN{5\textsuperscript{th} Roland Seebacher}
\IEEEauthorblockA{\textit{Electric Drives and}\\\textit{Power Electronic Systems Institute (EALS)}\\
\textit{Technische Universität Graz}\\
Graz, Austria}
\and
\IEEEauthorblockN{6\textsuperscript{th} Annette Mütze}
\IEEEauthorblockA{\textit{Electric Drives and}\\\textit{Power Electronic Systems Institute (EALS)}\\
\textit{Technische Universität Graz}\\
Graz, Austria}
\and
\IEEEauthorblockN{7\textsuperscript{th} Yvonne Späck-Leigsnering}
\IEEEauthorblockA{\textit{Institute for Accelerator Science}\\
\textit{ and Electromagnetic Fields (TEMF)}  \\
\textit{Graduate School CE}\\
\textit{Technische Universität Darmstadt}\\
Darmstadt, Germany}

}

\maketitle

\begin{abstract}
Accurate and efficient thermal simulations of induction machines are indispensable for detecting thermal hot spots and hence avoiding potential material failure in an early design stage. A goal is the better utilization of the machines with reduced safety margins due to a better knowledge of the critical conditions. In this work, the parameters of a two-dimensional induction machine model are calibrated according to evidence from measurements, by solving an inverse field problem. The set of parameters comprise material parameters as well as parameters that model three-dimensional effects. This allows a consideration of physical effects without explicit knowledge of its quantities. First, the accuracy of the approach is studied using an academic example in combination with synthetic data. Afterwards, it is successfully applied to a realistic induction machine model.
\end{abstract}


\section{Introduction}
\label{sec:intro}
%\cb{Add citation \cite{Loukrezis_2018}, \cite{De_Gersem_2020}}
The trends for higher utilization and increased power density and therewith heat generation, sparks an interest in the accurate thermal simulation of electric machines \cite{Soualmi_2018}, especially in highly constrained environments such as automotive applications. Numerous thin insulation layers of various materials \cite{Chapman_2008aa}, separating components on different electrical potentials, are prone to material failure due to overheating \cite{Pyrhonen_2013aa} or dielectric breakdown \cite{Hemmati.2019} which is a major cause of machine failure and limits the operating range. Their small thickness and their proximity to the windings, the main heat source, turn the particularly sensitive winding insulation into a component which necessitates thermal investigation \cite{Boglietti.2018}. Current thermal simulation is typically carried out with fast analytical circuit models based on lumped parameters or with the more complex and time-consuming numerical \ac{fe} method \cite{Boglietti_2009}. The latter is more capable of handling complex geometries and \acp{bc} \cite{Ejiofor_2019} and therefore more suited to analyse critical hot spots. Lumped parameter models are not capable of modelling localized effects, which are considered a critical aspect for diagnosing faults or failures.
Considering \ac{3d} models can be computationally demanding. Therefore, suitable approximations such as the reduction to \ac{2d} models are employed \cite{Ho.1998}. In this work, the temperature distribution is simulated in a \ac{2d} \ac{fe} model of a squirrel-cage induction machine. The \ac{2d} model is obtained by cutting the \ac{3d} model along its longitudinal cross-section. However, this simplified model omits relevant \ac{3d} effects, such as heat conduction in the axial direction and cooling by convection of the winding overhang \cite{Bramerdorfer.2018}. A pragmatic approach for compensating these shortcomings is manually fitting the \ac{2d} model parameters to measurement data \cite{Bergfried_2023aa}. The systematic introduction of data to enrich the model is a topic of ongoing research \cite{Kurz_2022}. Here, poorly known parameters such as, e.g., homogenized electric or thermal conductivities, convection coefficients or some local thermal heat sources are calibrated on the basis of measurement results along an inverse problem solving step \cite{Harrach.2021}. This approach is promising for a more precise identification of those poorly known parameters which are for example related to uncertainties in thermal material parameters, related to small and variable geometric details or related to variability in the heat sources.\\
This paper formulates an inverse thermal problem for model calibration.
%The thermal model and the inverse problem are formally introduced in Sections~\ref{sec:thermal} and \ref{sec:inverse} respectively. Afterwards, in Sec~\ref{sec:academic}, the proposed ...
In the first step, the proposed approach is validated on academic models. For this, synthetic measurement data are used.  %Subsequently, in Sec.~\ref{sec:machine}, ...
Subsequently, the inverse problem is transferred to the machine model, first using synthetic- and then real measurement data obtained from a lab machine at TU Graz. Comparison against an independent measurement set validates the model calibration approach. %The results are evaluated in Sec.~\ref{sec:results}, follwed by the conclusion of this paper.

%The results are compared against an independent measurement set to validate the approach, with the result, that the thermal parameter estimation of the induction machine is successful.
%By this calibration process, the accuracy of the forward simulation can be improved with respect to the real machine, better representing its thermal distribution. Relevant requirements and heuristic design rules can be derived. 
%After introducing the forward and inverse problem formulations, the simplified model is introduced and analysed. The estimation problem is then applied to the induction motor.

\section{Thermal Model}\label{sec:thermal}
This section introduces the thermal problem, its discretization in \ac{fe} analysis and homogenization as well as some remarks about \ac{3d} effects and homogenized model regions.

\subsection{The static heat conduction equation}
The static heat conduction equation describes the steady-state temperature distribution within the model domain $\Omega$: %\cite{Oziscik_89}:
\begin{subequations}
\begin{align}
	-\nabla\cdot\left( {\lambda\nabla{T}}\right) &= g \hspace{1cm} \text{in}\: \Omega \subset\mathds{R}^2, \,\label{eq:heatCond}\\
	T &= T_\text{0}  \hspace{0.8cm} \text{on}\: \Gamma_\text{D},\label{eq:iso}\\
	-\lambda\frac{\partial T}{\partial \mathbf{n}} &= 0 \hspace{1cm} \text{on}\: \Gamma_\text{N},
	 \label{eq:adi}\\
  \lambda\frac{\partial T}{\partial \mathbf{n}} + h(T-T_\text{0})&=0 \hspace{1cm} \text{on}\: \Gamma_\text{R}.
	 \label{eq:conv}
	 \end{align}%
	 \label{eq:heat_equation}%
	 \end{subequations}%
The solution, $T(\mathbf{x})$, is the temperature distribution, $T_0$ is the ambient temperature, $\lambda(\mathbf{x}, T)$ is the thermal conductivity, in general with spatial and temperature dependence, $h(\mathbf{x})$ is the heat-transfer coefficient and $g(\mathbf{x})$ is the internal heat source density term. The domain boundary is $\Gamma:=\partial \Omega$ and is decomposed into Dirichlet $\Gamma_\text{D}$, Neumann $\Gamma_\text{N}$ and Robin $\Gamma_\text{R}$ boundary parts. 
The Dirichlet \ac{bc} represents an isothermal coupling to the surrounding region. The Neumann \ac{bc} describes perfect thermal isolation on the boundary surface. The Robin \ac{bc} models thermal convection.\\
The field problem \eqref{eq:heat_equation} may include a set of parameters $\boldsymbol{\theta}$ that need calibration. Solving \eqref{eq:heat_equation} to obtain a solution $T(\mathbf{x}; \boldsymbol{\theta})$ constitutes the forward problem and is denoted by $F(\boldsymbol{\theta})$. The forward problem is solved numerically by means of the \ac{fe} method using the open-source solver \texttt{Pyrit} \cite{Bundschuh_2023ab}.

\subsection{Finite-element discretization}
The \ac{2d} cross-sectional model is meshed with triangles. The temperature 
\begin{equation}
    T \approx \sum_{j=1}^{N}u_j N_j \label{eq:FE_temp}
\end{equation}
is discretized by first-order nodal shape functions $N_j(\mathbf{x})$ and degrees of freedom $u_j (\boldsymbol{\theta})$.\\
The Ritz-Galerkin \cite{Fletcher.1984} procedure leads to the system of equations
\begin{equation}
    \mathbf{K}_\lambda \mathbf{u} = {\mathbf{g}} \label{eq:weak_heat},
\end{equation}
with the stiffness matrix and load vector
\begin{subequations}
    \begin{align}
        K_{\lambda,ij} &= \int_\Omega \lambda \nabla N_j \cdot \nabla N_i \, d\Omega \\
        g_i &= \int_\Omega g \, N_i \, d\Omega .
    \end{align}
    \label{eq:fem_kq}
\end{subequations}  

\subsection{3D effects and homogenization}
%%\cb{To reduce model complexity and computation times, the electric machine is simulated as a \ac{2d} model \cite{Belahcen.2010}. This reduction introduces systematic errors. However, by determining equivalent parameters, this systematic error is significantly reduced. These equivalent parameters account for 3D effects such as axial conduction or convective effects, which would otherwise be neglected. In the electrical machine these effects are present in the anisotropic lamination of iron in stator and rotor or in the heat flux in the winding overhang and air gap \cite{Staton_2005}. Another reduction of complexity is the homogenization of small geometric regions with similar thermal properties as the insulation system in the electric machine.\\The equivalent parameters are determined from measurements of the real machine in the inverse problem.}\\
To reduce model complexity and computation times, a \ac{2d} model is simulated \cite{Marfoli_2017}. This model entails systematic errors that require consideration. Errors occur due to the neglect of \ac{3d} effects such as conduction in axial direction. Furthermore, thermal convection effects at the front and back machine side are not modelled \cite{Staton_2005}. Another approximation is introduced by inherently homogenizing small regions with similar thermal properties, which averages small-scale effects. \cite{Wang_2006}. These systematic errors are significantly reduced by introducing equivalent model parameters that are calibrated according to measurement data in the inverse problem.

%This reduction introduces systematic errors due to the neglect of \ac{3d} effects due to axial conduction. 

%These are due to axial conduction in the laminated stator and rotor iron, in the rotor shaft and bearing and due to convective effects in the winding overhang, the end shields and the air gap \cite{Staton_2005}.
%Furthermore, composite layers, that represent multiple regions with similar thermal properties and/or with small geometric dimensions are introduced. These require an appropriate approximation of the heterogeneous thermal properties and interface effects \cite{Wang_2006}.\\
%This paper presents an approach to approximate these effects by defining equivalent parameters. Their values are determined from measurements of the real machine in the inverse problem.

\section{Inverse Problem}\label{sec:inverse}
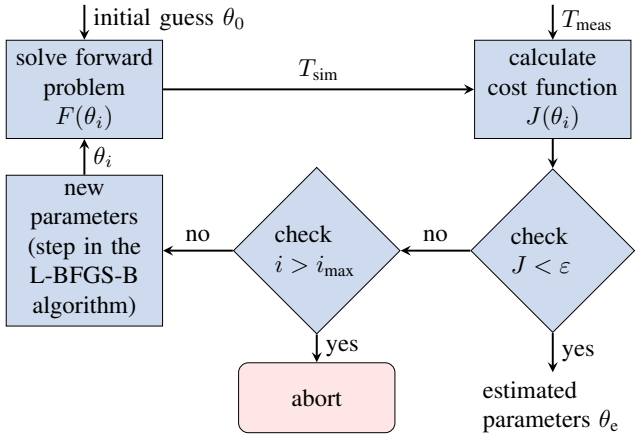
\begin{figure}
    \centering
    \resizebox{1.1\linewidth}{!}{
    \begin{tikzpicture}[
    abort/.style = {rectangle, rounded corners, text width=2cm, minimum height=1cm,text centered, draw=black, fill=red!10, align=center, anchor=center},
    box/.style = {rectangle, text width=2cm, minimum height=1cm, text centered, draw=black, fill=tikzblau!30, align=center, anchor=center},
    decision/.style = {diamond, text width=1.2cm, minimum height=1cm, text centered, draw=black, fill=tikzblau!30, align=left, anchor=center},
    arrow/.style = {thick,->,>=stealth}
]
    \matrix (m) [matrix of nodes, row sep=0.2cm, column sep=1cm, text width=2cm, align=left] {
        |[box]|  solve forward problem $F(\theta_i)$ &  & |[box]|  calculate cost function $J(\theta_i)$\\
        & & & \\
        |[box]| new parameters (step in the L-BFGS-B algorithm) & |[decision]| check $i > i_\text{max}$ & |[decision]| check $J<\varepsilon$ \\
        & & & \\
        & |[abort]| abort  & estimated parameters $\theta_\text{e}$ \\
    };

    \draw[arrow] (m-1-1) edge node[above] {$T_\text{sim}$} (m-1-3);
    \draw[arrow] (m-1-3) edge (m-3-3);
    \draw[arrow] (m-3-3) edge node[above] {no} (m-3-2);
    \draw[arrow] (m-3-2) edge node[above] {no} (m-3-1);
    \draw[arrow] (m-3-2) edge node[right] {yes} (m-5-2);
    \draw[arrow] (m-3-1) edge node[right] {$\theta_i$} (m-1-1);
    \draw[arrow] ([yshift=0.5cm]m-1-1.north) -- (m-1-1.north) node[above right] {initial guess $\theta_0$};
    \draw[arrow] (m-3-3) edge node[right] {yes} (m-5-3);
     \draw[arrow] ([yshift=0.5cm]m-1-3.north) -- (m-1-3.north) node[above right] {$T_\text{meas}$};
\end{tikzpicture}}
    \caption{Inverse parameter estimation process. Beginning with an initial guess $\theta_0$, the estimated parameters $\theta_\text{e}$ are obtained in an iterative process.}
    \label{fig:inverse}
\end{figure}

The inverse problem reads
\begin{equation}
    \begin{cases}
        \underset{\boldsymbol{\theta}}{\min}\,J(\boldsymbol{\theta}),\\
        \text{such that \eqref{eq:heat_equation} is fulfilled.}
    \end{cases}
\end{equation}
Here,
\begin{equation}
    J(\boldsymbol{\theta}) = \sum_{i=1}^{N_{\text{samples}}} \sum_{j=1}^{N_\text{sensors}} (T_{\text{sim},i}(\mathbf{x}_j; \boldsymbol{\theta})-T_{\text{meas},i,j})^2 \label{eq:loss_func}
\end{equation}
is the cost function. $T_{\text{sim},i}(\mathbf{x}_j; \boldsymbol{\theta})$ are temperatures obtained by evaluating the solution of the parameterized forward problem $F(\boldsymbol{\theta})$ evaluated at the sensor position $x_j$ and $T_{\text{meas},i,j}$ are the temperatures measured by the sensors. Index $j = 1,...,N_\text{sensor}$ and index $i = 1,...,N_\text{samples}$ count the sensors and the sampled operating points respectively. Therefore, \eqref{eq:loss_func} compares simulated and measured temperatures at discrete positions for various operating conditions.\\
Figure \ref{fig:inverse} outlines the iterative inverse parameter estimation process. The iteration is finished successfully after falling below a prescribed tolerance $\varepsilon = 10^{-6}$ or otherwise, is aborted if a maximum number of iterations is reached. The optimization method used in this study is the limited-memory-Broyden-Fletcher-Goldfarb-Shanno-bounded (L-BFGS-B) algorithm, as it is a standard optimizer available in scientific toolboxes. This algorithm is a bounded quasi-Newton algorithm using limited memory to approximate and update gradient information \cite{Byrd_1995} to calculate the new set of parameters for each iteration (bottom left of Fig. \ref{fig:inverse}) until the terminal criteria is met. 
Other optimization methods can be used as well.

\section{Academic Examples}\label{sec:academic}
\begin{figure}
    \centering
    \includegraphics[width=.8\linewidth]{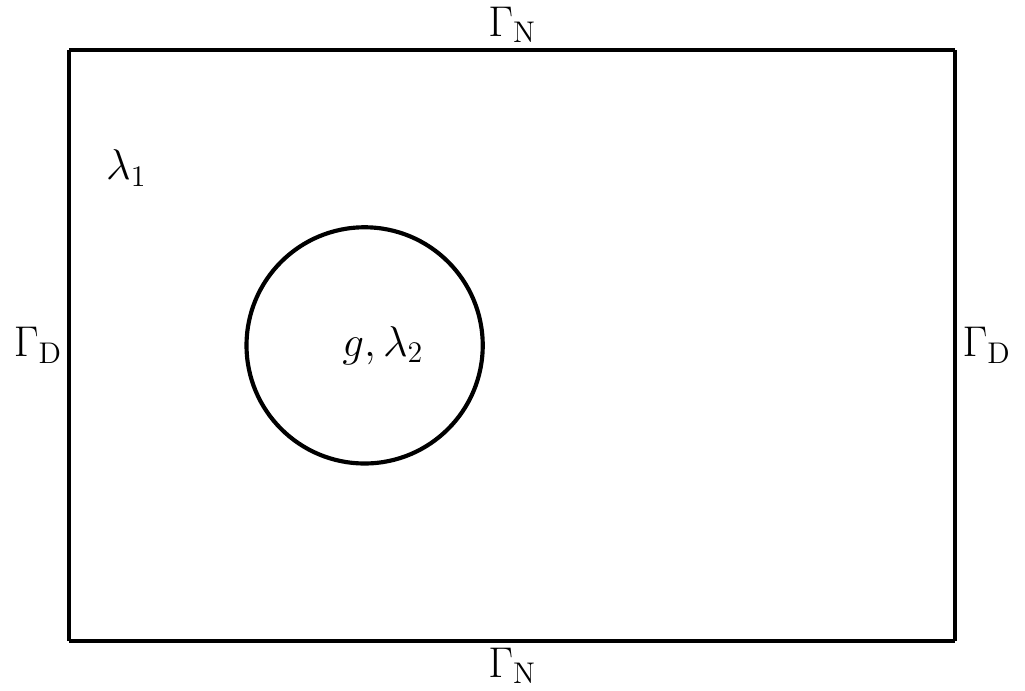}
    \caption{Geometry of example 1. The model consists of a domain with thermal conductivity $\lambda_1$ and a domain with generated heat $g$ and thermal conductivity $\lambda_2$.}
    \label{fig:academic_1}
\end{figure}

\begin{figure}
    \centering
    \includegraphics[width=.8\linewidth]{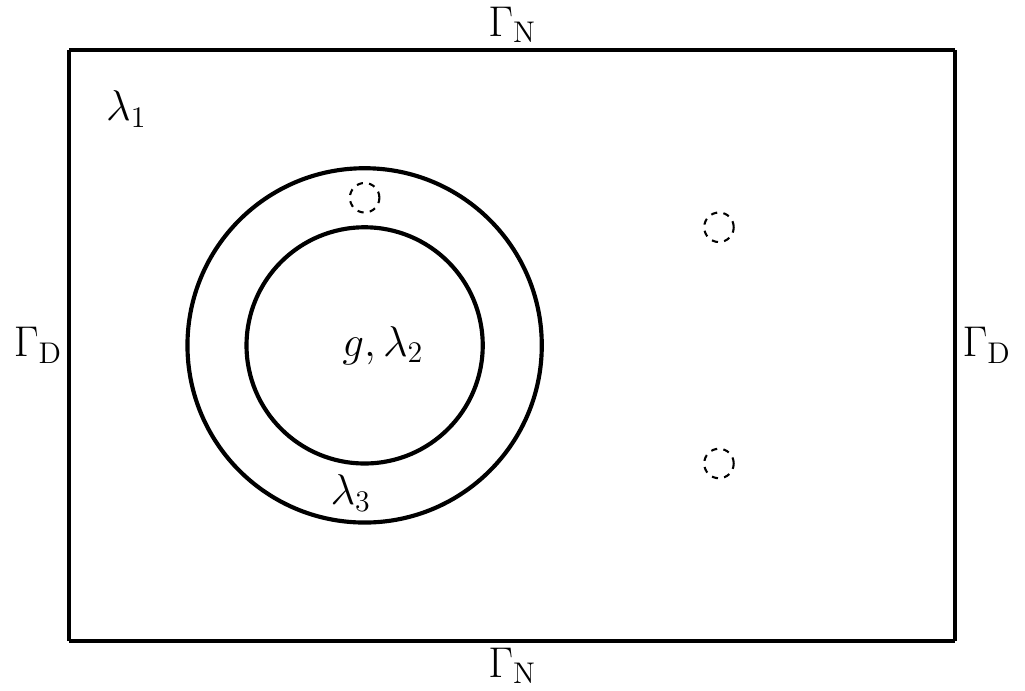}
    \caption{Geometry of example 2 with discrete sensors, visualized by dotted circles. In addition to example 1, a cylindrical layer with thermal conductivity $\lambda_3$ is added.}
    \label{fig:academic_2}
\end{figure}

At first, the inverse parameter estimation is applied to two academic examples. 
The geometry of example 1 is depicted in Fig.~\ref{fig:academic_1}. It consists of a rectangle with a combination of isothermal and adiabatic boundaries $\Gamma_\text{D}$ and $\Gamma_\text{N}$. Additionally, a circular heat source is modelled. This example is used to investigate the solvability of the inverse problem and the  accuracy of its solution. 
The geometry of example 2 is depicted in Fig.~\ref{fig:academic_2}. In comparison to the first example, an additional domain surrounding the heat source and a number of discrete sensors are added. The additional domain models an insulation layer surrounding the heat source (i.e. winding). This example is used to investigate the sensitivity of the inverse problem with respect to the available temperature information.

For both examples, the thermal problem is solved with a parameter set $\boldsymbol{\theta} = [\lath{1}, \lath{2}, \lath{3}]$. Synthetic measurement data are generated by solving the forward problem with the ground-truth values and afterwards perturbing the solution with noise. The noise is a standard distributed additive error with the standard deviation $\sigma^2$. The noise is chosen to $\sigma^2$ = 1 K in accordance to the expected measurement noise. For example 1, these data are generated for each node of the \ac{fe} mesh, whereas for example 2, only data for the nodes at the sensor positions are generated. 
To evaluate the accuracy of the solution of the inverse problem, the relative error is defined as 
\begin{equation}
\varepsilon_{\text{rel}, i} =  \frac{\lvert\lambda_{i, \text{estimated}} - \lambda_{i, \text{true}}\rvert}{\lambda_{i, \text{true}}} .
    \label{eq:relErr}
\end{equation}

\subsection{Example 1: Rectangle with heat source}
\begin{figure}
    \centering
    \begin{tikzpicture}
		\centering
		\begin{axis}[scale=0.8,grid=major,
		             width=4in,
		             height=2in,
		             xmode=log,
		             ymode=log,
		             xlabel={Number of independently perturbed simulations},
		             ylabel={Relative error $\varepsilon_{\text{rel}}$}]

	    \addplot[mark=*, mark options={solid}, thick, black] 
	  table[row sep=crcr]{%
   1.000000000000000000e+00 7.365351520800415613e-04\\
5.000000000000000000e+00 3.774890516199529990e-04\\
1.000000000000000000e+01 2.609794287904584854e-04\\
5.000000000000000000e+01 9.740666956990822129e-05\\
1.000000000000000000e+02 7.507652967076603931e-05\\
5.000000000000000000e+02 3.546027706560411162e-05\\
1.000000000000000000e+03 2.239143273328837372e-05\\
5.000000000000000000e+03 1.297876602647626835e-05\\
};
    \addlegendentry{$\varepsilon_{\text{rel},1}$}
        \addplot[mark=*, mark options={solid}, thick, blue] 
	  table[row sep=crcr]{%
   1.000000000000000000e+00 4.418584099850594079e-04\\
5.000000000000000000e+00 2.152931302736786178e-04\\
1.000000000000000000e+01 1.913457767349561578e-04\\
5.000000000000000000e+01 6.613990716869041218e-05\\
1.000000000000000000e+02 4.288601696995630603e-05\\
5.000000000000000000e+02 2.129573824863525007e-05\\
1.000000000000000000e+03 1.065747171006640495e-05\\
5.000000000000000000e+03 5.824285123109026488e-06\\
   };
    \addlegendentry{$\varepsilon_{\text{rel},2}$}

    %\addplot[mark=none] coordinates {(1, 9.74e-06) (5, 9.74e-06) (1, 2.61e-05) (1, 9.74e-06)};
    \addplot[mark=none] coordinates {(10, 9.74e-06) (100, 9.74e-06) (10, 3.775e-05) (10, 9.74e-06)};
		\end{axis}
	\draw[](2.8, 0.09) node[above, text=black]{1};
	\draw[](1.8, 0.5) node[above, text=black]{0.5};
	\end{tikzpicture}
    \caption{Relative error of example 1 over number of independently perturbed simulation with multiple seeds.}
    \label{fig:Conv_Model2}
\end{figure}
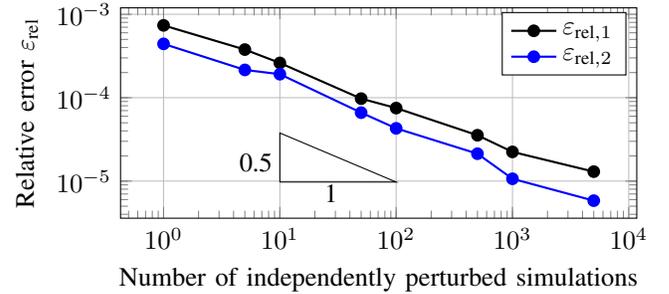

Figure~\ref{fig:Conv_Model2} shows the relative errors $\varepsilon_{\text{rel},i}$ for an increasing number of independently perturbed simulations. The relative error of the estimation with 10 simulations ($N=10$) is below 0.1\,\% and decreases below 0.002\,\% for $N=10^3$. 
Therefore, the parameter estimation of example 1 is conducted successfully.

\subsection{Example 2: Rectangle with insulated heat source}
\begin{figure}
    \centering
    \begin{tikzpicture}
    \centering
    \begin{axis}[scale=0.8,grid=major,
                 width=4in,
	             height=2in,
                 xmode = log,
                 ymode = log,
                 xlabel={Number of independently perturbed simulations},
                 ylabel={Relative error $\varepsilon_{\text{rel}}$}]

\addplot[mark=*, mark options={solid}, thick, black] 
	  table[row sep=crcr]{%
   1.000000000000000000e+00 2.950620412228022738e-01\\
5.000000000000000000e+00 8.311533489405213693e-02\\
1.000000000000000000e+01 7.663111999151880904e-02\\
5.000000000000000000e+01 2.889028376940486198e-02\\
1.000000000000000000e+02 2.604474385476883078e-02\\
5.000000000000000000e+02 1.157223611244557332e-02\\
1.000000000000000000e+03 8.724992583087499282e-03\\
5.000000000000000000e+03 4.290937757546753502e-03\\
1.000000000000000000e+04 2.936370588121910759e-03\\
   };
    \addlegendentry{$\varepsilon_{\text{rel},1}$};
\addplot[mark=*, mark options={solid}, thick, blue]
	  table[row sep=crcr]{%
   1.000000000000000000e+00 4.257699467066117932e-02\\
5.000000000000000000e+00 1.334268519767258313e-02\\
1.000000000000000000e+01 1.024954181281166957e-02\\
5.000000000000000000e+01 5.413728477799845291e-03\\
1.000000000000000000e+02 3.977131820287837724e-03\\
5.000000000000000000e+02 2.109668458286578698e-03\\
1.000000000000000000e+03 1.594813684361817696e-03\\
5.000000000000000000e+03 6.156905497357145329e-04\\
1.000000000000000000e+04 4.693711667543787542e-04\\
   };
\addlegendentry{$\varepsilon_{\text{rel},3}$}
 %\addplot[mark=none] coordinates {(10, 3.977e-03) (100, 3.977e-03) (10, 1.025e-02) (10, 3.977e-03)};
\addplot[mark=none] coordinates {(10, 9.725e-04) (100, 9.725e-04) (10, 2.7044e-03) (10, 9.725e-04)};
    \end{axis}
\draw[](2.7, 0.09) node[above, text=black]{1};
\draw[](1.7, 0.5) node[above, text=black]{0.3};
\end{tikzpicture}
    \caption{Relative error of example 2 over number of independently perturbed simulations.}
    \label{fig:Conv_Model6}
\end{figure}
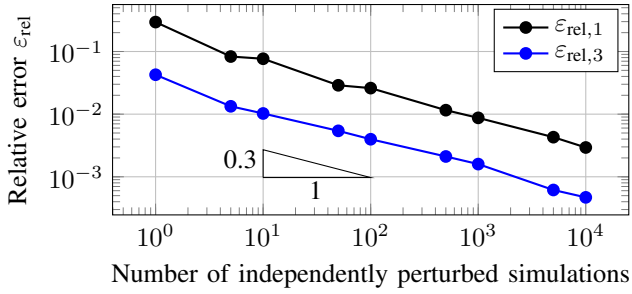

Example 1 proved the solvablity of the inverse problem and accuracy of its solution, when measurement data are available on each node of the \ac{fe} mesh. However, in reality the temperatures are only known at the locations of the sensors. Therefore, for example 2, the inverse problem is solved using only temperature information of the nodes associated with the discrete sensors. In this case, only 3 sensors are considered.
The relative error is 1\,\% for 10 measurement samples and below 0.1\,\% for  $N > 10^3$ and converges as shown in Fig.~\ref{fig:Conv_Model6}.\\
Example 2 already reflects the nature of the upcoming technical example, i.e., the domains $\Omega_1$, $\Omega_2$ and $\Omega_3$ mimic the iron stator core, copper coil and epoxy resin insulation of the induction machine respectively. In the insulation layer, the sensitivity of the temperature to $\lambda$ given as $\frac{\mathrm{d}T}{\mathrm{d}\lambda_i}(\mathbf{x})$ is maximal because $\lambda_3 \ll \lambda_{1}$ and $\lambda_3 \ll \lambda_{2}$. Therefore, a sensor is required in this region to evaluate this sensitivity and improve the parameter estimation of $\lambda_3$. The resulting relative errors are small compared to the accuracy of the used sensors in the machine, the parameter estimation of the second model is also conducted successfully.

\section{Induction machine}\label{sec:machine}
\subsection{Machine data}
\begin{figure}
    \centering
    \scalebox{0.55}{\input{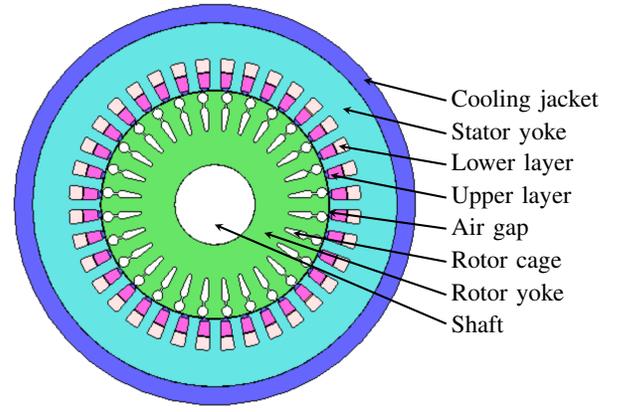}}
    \caption{Cross section of the induction machine. The stator and rotor yoke is laminated iron, the lower and upper layer combines the insulation system, which is epoxy resin and the windings, which are copper, the rotor cage is aluminium and the shaft is steel \cite{Bergfried_2023aa}.}
    \label{fig:machine_names}
\end{figure}

\begin{figure}
    \centering
    \scalebox{0.55}{\input{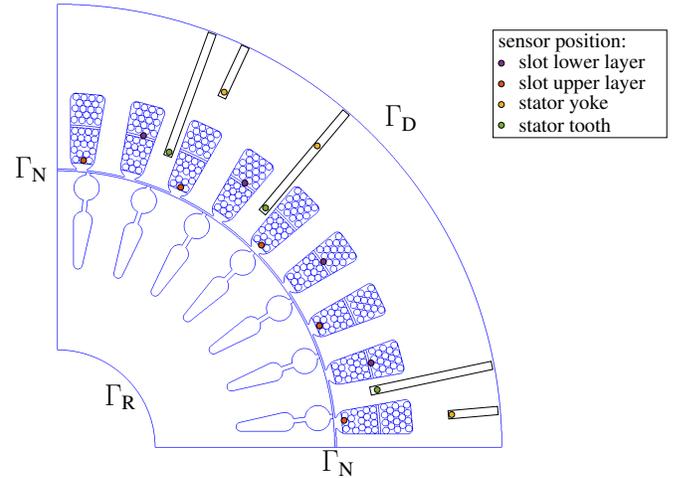}}
    \caption{Machine model including the position of the sensors, the applied boundary conditions and the resolved windings in the slot.}
    \label{fig:pos_sensors}
\end{figure}

Example 3 deals with a four-pole squirrel-cage induction machine \cite{Eickhoff_2020aa} with a rated power of 3.7~kW at 1430 rpm. The 36 slots are filled with two-layer wire windings immersed in insulating epoxy resin. A sketch of the machine and its different components is given in Fig. \ref{fig:machine_names}.  
The temperature sensors are buried in the stator iron and the stator slots as shown in Fig. \ref{fig:pos_sensors}. For measurement purposes the machine is encased in a cooling jacket. The machine can be operated at various drive cycles \cite{Heidarikani_2023} at a controllable temperature.
\subsection{Measurement procedure}
\begin{figure}
    \centering
    \scalebox{0.65}{\input{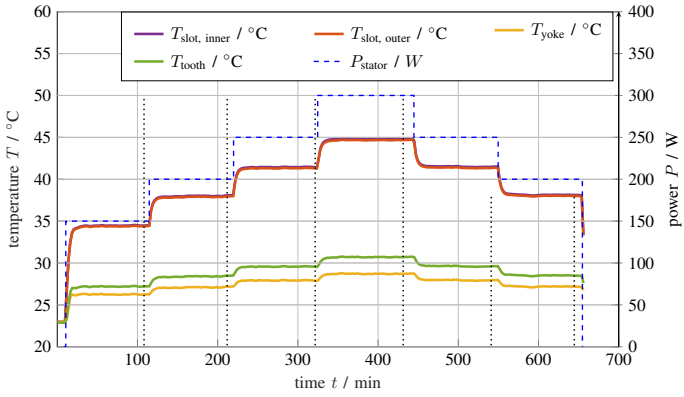}}
    \caption{Measured temperatures in the sensors (solid). The stator power is drawn in blue (dashed). The dotted vertical lines indicate six times, at which a steady-state is assumed and the samples are taken.}
    \label{fig:measurement}
\end{figure}

In the experimental setup, a constant electric power $P$ is applied to the series connection of all windings of the machine at standstill. The resulting volumetric heat generated in the winding is 
\begin{equation}
    g = \frac{P}{V},
\end{equation}
where $V$ is the volume of the copper winding. The heat source is generated due to Joule losses in the winding and depends on the winding resistance and current.
%\lb{When modelling the winding as a boundary which introduces heat to the surrounding regions, the heat flux across the winding surface area is considered. In general, the heat source in \eqref{eq:heatCond} is a volumetric quantitiy\cite{Hahn.2012}.}
During measurement, $P$ or the cooling jacket temperature $T_0$ are increased in steps. A temperature sample is taken once a steady-state is observed from a limited change in temperature over time. A measurement run is shown in Fig. \ref{fig:measurement}, where $P$ is varied in steps of 50~W, starting from 150~W, whereas $T_0$ is fixed to 25\,$^\circ$C. From this measurement run, six steady-state samples at four operation points are obtained. 
The dotted vertical lines indicate the points at which a steady-state is observed and a sample is taken. A sample contains measurement data from all sensors.

In total, measurement data are comprised of 20 samples at six different operating points. The operating points combine $P$ between 150\,W and 300\,W and $T_0$ between 20\,$^\circ$C and 35\,$^\circ$C. Thanks to azimuthal symmetries, sensor data of the whole machine can be utilized, while only considering a quarter of the machine. Therefore, measurement data of 80 samples become available, which are split into two sets, 48 used for calibration and 32 used for independent validation. Either set contains a broad selection of operating points. The calibration set is richer in order to achieve an accurate calibration.
%The result of the calibration is a set of parameters that can be used in the forward simulation to generate a calibrated temperature distribution for a given operating point. This distribution can then be compared to the initial forward simulation against the validation data set.

\subsection{Parameters selected for calibration}\label{sec:results}
The parameters to be calibrated in the machine are $\boldsymbol{\theta}=[\lambda_\text{air},\lambda_\text{insulation},\lambda_\text{stator iron},\lambda_\text{rotor iron}]$. $\lambda_\text{air}$ models conduction, convection as well as radiation effects in the air gap. $\lambda_\text{insulation}$ models the homogenized insulation consisting of several insulation layers such as winding enamels, layer separation and slot liners, as well as epoxy resin inclusions. Representing each layer individually raises the complexity of the model significantly due to different geometric scales by multiple orders of magnitude. A homogenization of these regions is expected to reduce the model complexity and speed up the computations, but the accuracy of the results needs to be verified. Exact material and homogenized properties are not known and are therefore estimated from experimental data.
$\lambda_\text{stator iron}$ and $\lambda_\text{rotor iron}$ model thermal conduction through the stator and rotor lamination stacks.\\
In example 3, the results are calculated for a model that uses synthetic measurement data. Here, the quality of the estimation can be accurately assessed by comparing it to the available ground-truth values. In example 4, the actual measurement data are used on the model. The confidence in the estimation result can be trusted based on the previously learned estimation accuracy, but measurement and systematic uncertainties have to be kept in mind. The calibrated system is then compared to a set of validation measurements.

\subsection{Example 3: Induction machine model with synthetic measurement data}
The results for the inverse machine problem are shown in Tab. \ref{tab1} for $N = 10$ and $N = 10^3$ samples of the synthetic measurements data. The principle of data generation is similar to the first examples, where the forward solution is calculated with ground-truth data. This solution is then perturbed $N$ times independently. As in the example 1, the synthetic data for the whole machine are used for the inverse estimation. A good estimation is achieved for all parameters except for $\lambda_\text{rotor iron}$. The measurements are insensitive to $\lambda_\text{rotor iron}$ because of the large thermal resistance of the air gap and because of the fact that no losses are introduced in the rotor. The conductivity is furthermore assumed to be equal to $\lambda_\text{stator iron}$ because the same material is used.\\
Next, similar to example 2, only synthetic data evaluated at the sensor positions are used. Here, a deterioration of the estimation of $\lambda_\text{air}$ can be observed. A variation of its (already small) value can not be detected by the sensors.

\renewcommand{\arraystretch}{1.2}
\begin{table}[htbp]
\caption{Results of the inverse problem with synthetic measurements}
\begin{center}
\begin{tabular}{@{}clrr@{}}%{|c|l|r|r|}
\hline
\textbf{Used}&\multicolumn{3}{c}{\textbf{Estimation results}} \\
%\cline{2-4} 
\textbf{Data} & \textbf{Parameter}& \textbf{\textit{$\varepsilon_\text{rel}$ ($N=10$)}}& \textbf{\textit{$\varepsilon_\text{rel}$ ($N=1000$)}} \\
\hline
synthetic & $\lambda_\text{air}$ & 2\,\% & 2\,\% \\
measurements & $\lambda_\text{insulation}$ & 0.5\,\% & 0.25\,\% \\
without & $\lambda_\text{stator iron}$ & 0.07\,\% & 0.04\,\% \\
sensors & $\lambda_\text{rotor iron}$ & 50\,\% & 50\,\% \\
\hline
synthetic & $\lambda_\text{air}$ & 100\,\% & 10\,\% \\
measurements & $\lambda_\text{insulation}$ & 2\,\% & 0.1\,\% \\
with & $\lambda_\text{stator iron}$ & 0.9\,\% & 0.03\,\% \\
sensors & $\lambda_\text{rotor iron}$ & 100\,\% & 200\,\% \\
\hline
\end{tabular}
\label{tab1}
\end{center}
\end{table}

\subsection{Example 4: Induction machine model with real measurement data}
Table \ref{tab2} lists the estimated parameters using all available calibration data ($N=48$).
According to the experience gained from example 3, $\lambda_\text{rotor iron}$ is no longer selected for calibration.\\
Temperature gradients in the cooling jacket are observed as varying values for the nearby $\lambda_\text{stator iron}$ when simulating different sections of the machine. These differences are attributed to the good thermal conductivity of iron and the not perfectly symmetrical sensor positions. This influence is filtered out by averaging along the full circumference of the motor.

\begin{table}[htbp]
\caption{Results of the inverse problem using real measurement data.}
\begin{center}
\begin{tabular}{lc}
\hline
\multicolumn{2}{c}{\textbf{Estimation results}} \\
\textbf{Parameter}& $\lambda$ [$\frac{\text{W}}{\text{mK}}$] for \textbf{\textit{$N=48$}} \\
\hline
$\lambda_\text{air}$ &  0.037\\
$\lambda_\text{insulation}$  & 0.073\\
$\lambda_\text{stator iron}$ &  30.70\\
\hline
\end{tabular}
\label{tab2}
\end{center}
\end{table}
The error of the found solution is evaluated by comparing the results of a forward problem $T_\text{cal}$ which uses the parameters of Table \ref{tab2} with the validation set $T_\text{val}$. The relative error $\varepsilon_{\text{rel},T}$ is defined by
\begin{equation}
\varepsilon_{\text{rel,T},i} =  \frac{\lvert T_{\text{cal},i} - T_{\text{val},i}\rvert}{T_{\text{val},i}} .
    \label{eq:relErr2}
\end{equation}
Averaging over all available validation data, relative errors of 0.64\,\% for the stator yoke, 0.34\,\% for the stator teeth and 0.77\,\% for the stator slots are obtained. Additionally, the relative error for all investigated operation points is below the measurement accuracy of the used temperature sensors. Therefore, the calibration is considered successful and a good thermal model of the machine is acquired.

\section{Conclusion}\label{sec:conclusion}
\ac{2d} \ac{fe} models allow an efficient way to thermally characterize induction machines and predict hot spots. Neglecting \ac{3d} effects would lead to a poor accuracy. Here, \ac{3d} effects are inserted in the \ac{2d} model by calibrating a few of the model parameters according to available measurement data. The calibration is carried out by solving an inverse thermal field problem, enforcing the simulated temperature to be close to the temperatures measured by a set of sensors.\\
The paper successfully validates the approach for two academic examples and a third example being an induction machine model with synthetic measurement data. In a fourth example, it was shown that the method is capable of calibrating a \ac{2d} \ac{fe} thermal induction machine model from measurement results, thereby including \ac{3d} effects.

%\ac{2d} \ac{fe} models are an efficient way to simulate induction machines and to detect potential hot spots. To obtain an accurate solution, \ac{3d} effects have to be considered. This is achieved with the \ac{2d} model by calibrating some of the model parameters.\\
%It has been shown that the thermal inverse problem is a systematic approach to determine these parameters. In academic examples using synthetic measurements, the accuracy of the solution was investigated. Afterwards, the inverse problem was applied to the induction machine model, first with synthetic measurements and then with real measurements. The calibration via inverse problem shows to be a promising approach for the determination of the thermal parameters in the stator and slot area. The resulting calibrated machine model shows good accordance of the simulated temperatures with the measured temperatures.

%\cb{@YSL @HDG: In anderen ICEM Papern wurde dieser Abschnitt Conclusion and Future Work genannt. Es wäre also eine Möglichkeit hier viele Creator-Kooperationsverweisen einzufügen. Mit Referenzen zu anderen Papern. Meinung???}

\section*{Acknowledgment}
We like to thank H. Eickhoff for designing and realising the induction machine for measurements, J. Bundschuh for the support with Pyrit, as well as D. Loukrezis for the fruitful discussions. L. Blumrich likes to thank the Erasmus+ programme of the EU for supporting his research stay at TU Graz.%providing a mobility grant for the exchange with Graz.

%\section*{Reference}
%\bibliographystyle{plain}
\bibliographystyle{unsrt}
\bibliography{Bibfile_ICEM.bib} 

%\textcolor{red}{YSL: Folgende Punkte zu Bios:
%\begin{itemize}
%    \item Klären, ob die Bios bereits bei erster Einreichung in den Text sollen. Aus meiner Sicht ist das bei der ersten Einreichung unüblich. Oft wird das erst nach der Acceptance eingefügt. 
%    \item Bios homogenisieren. Ich würde bei Erstautor Technische Universität Darmstadt (TUDa) einführen und das bei den anderen dann als eingeführt annehmen. Das schafft Platz. ich vermeide "Technical University" um verwechslung mit FH zu vermeiden. 
%    \item TEMF analog einführen (und nicht bei Jeder bio wieder neu definieren.)
%\end{itemize}}

%\begin{IEEEbiography}[{\includegraphics[width=1in,height=1.25in,clip,keepaspectratio]{figs/Blumrich_Leon_sw_2.jpg}}]{Leon Blumrich} 
\begin{IEEEbiographynophoto}{Leon Blumrich}
received the B.Sc. and
M.Sc. degrees in mechatronics engineering, with a focus on mechatronic drives, from the \ac{tuda}, in 2021 and 2023, respectively. Starting in 2024, he is currently pursuing a Ph.D. degree on design and optimization of three-phase electrical machines at the Institute of Electrical Drive Systems (EAS) of \ac{tuda}. His research interests include multi-physical simulation and multi-objective optimization of permanent magnet synchronous machines in the context of automotive applications.
\end{IEEEbiographynophoto}

%\begin{IEEEbiography}[{\includegraphics[width=1in,height=1.25in,clip,keepaspectratio]{figs/Bergfried_temf_sww.jpg}}]{Christian Bergfried}
\begin{IEEEbiographynophoto}{Christian Bergfried}
     received the B.Sc. and
M.Sc. degrees in computational engineering from the \ac{tuda}, in 2019 and 2021, respectively, where he is currently pursuing a Ph.D. degree on finite-element simulations of electric and thermal stresses in electric machines at the \ac{temf} of \ac{tuda}. His research interests include electrothermal modeling and simulation of electric machines, material aging and detection of partial discharges.
\end{IEEEbiographynophoto}

%\begin{IEEEbiography}[{\includegraphics[width=1in,height=1.25in,clip,keepaspectratio]{figs/Armin_G_SW.jpeg}}]{Armin Galetzka}
\begin{IEEEbiographynophoto}{Armin Galetzka}
     received the M.Sc. degree in electrical engineering and information technology, along with the Dr.-Ing. degree from the \ac{tuda}, Darmstadt, Germany, in 2018 and 2023, respectively. As part of his doctoral research, he investigated methods for calculating electromagnetic fields, taking into account data-driven material models, as well as surrogate modeling and uncertainty quantification. In 2023, he joined the Carl Zeiss SMT GmbH.
\end{IEEEbiographynophoto}

%\begin{IEEEbiography}[{\includegraphics[width=1in,height=1.25in,clip,keepaspectratio]{figs/bio_degersem.png}}]{Herbert De Gersem} (Member, IEEE)
\begin{IEEEbiographynophoto}{Herbert De Gersem} (Member, IEEE)
    received the M.Sc. and Ph.D. degrees in electrical engineering from the KU Leuven, Leuven, Belgium, in 1994 and 2001, respectively. From 2001 to 2006, he was with the \ac{tuda}, Darmstadt, Germany. Since 2001, he has been an Associated Professor with the KU Leuven, Leuven, Belgium. Since 2014, he has been the Full Professor and the Head of the \ac{temf}, at the \ac{tuda}, Darmstadt, Germany. His research interests include finite-element electromagnetic field simulation for electrotechnical devices and particle accelerators and FDTD and FETD techniques for electromagnetic and ultrasonic wave propagation.
\end{IEEEbiographynophoto}

\begin{IEEEbiographynophoto}{Roland Seebacher}
received the Dipl.-Ing. and Doctoral degrees in electrical engineering from the TU Graz, Graz, Austria, in 1991 and 1996, respectively. He is currently an Assistant Professor with the Electric Drives and Machines Institute, Graz University of Technology. His research interests include modeling and control of electric machines and drives.
\end{IEEEbiographynophoto}

\begin{IEEEbiographynophoto}{Annette Mütze} (Fellow, IEEE)
    received the Ph.D. degree in electrical engineering from the TU Darmstadt, Darmstadt, Germany, in 2004. After working as an assistant professor, and later associate professor, at the universities of Wisonconsin-Madison, USA (until 2007) and Warwick, School of Engineering (until 2010), she became a university professor at the Electric Drives and Power Electronic Systems (EALS) institute, TU Graz, Austria, in 2010. Her main research areas are the improvement of reliability, efficiency and utilization of variable speed electrical drives, with substantial publications investigating the influence of bearing currents, manufacturing, control- and drive topologies and fractional horsepower drives.
\end{IEEEbiographynophoto}

%\begin{IEEEbiography}[{\includegraphics[width=1in,height=1.25in,clip,keepaspectratio]{figs/yvonne_temf_sww.jpg}}]{Yvonne Späck-Leigsnering}
\begin{IEEEbiographynophoto}{Yvonne Späck-Leigsnering}
received the M.Sc.\ degree in electrical engineering and information technology, along with the Dr.-Ing.\ degree from the \ac{tuda}, in 2014 and 2019, respectively. During her PostDoc tenure with the \ac{temf}, she founded the Quasistatics in Computational Engineering (QuinCE) research group, which she has been leading independently as a Athene Young Investigator Fellow since 2022. In 2023, she joined Robert Bosch GmbH. Ms. Späck focuses on electromagnetic modeling and simulation for electric equipment in power engineering and automotive applications. Her contributions were recognized with the Young Scientist of the Year award from the Werner-von-Siemens-Ring Foundation in 2023 and the Fritz und Margot Faudi Foundation award in 2023.
\end{IEEEbiographynophoto}

\begin{acronym}
	\acro{2d}[2D]{two-dimensional}
	\acro{3d}[3D]{three-dimensional}
	\acro{ac}[AC]{alternating current}
    \acro{bc}[BC]{boundary condition}
	\acro{dc}[DC]{direct current}
	\acro{dof}[DoF]{degree of freedom}
	\acroplural{dof}[DoFs]{degrees of freedom}
	\acro{em}[EM]{electromagnetic}
	\acro{epdm}[EPDM]{ethylene propylene diene monomer rubber}
	\acro{eqs}[EQS]{electroquasistatic}
	\acro{eqst}[EQST]{electroquasistatic-thermal}
	\acro{es}[ES]{electrostatic}
	\acro{fe}[FE]{finite-element}
	\acro{fem}[FEM]{finite-element method}
	\acro{fgm}[FGM]{field grading material}
	\acro{hv}[HV]{high-voltage}
	\acro{hvac}[HVAC]{high-voltage alternating current}
	\acro{hvdc}[HVDC]{high-voltage direct current}
	\acro{lsr}[LSR]{liquid silicone rubber}
	\acro{sir}[SiR]{silicone rubber}
	\acro{mo}[MO]{metal oxide}
	\acro{pd}[PD]{partial discharges}
	\acro{pde}[PDE]{partial differential equation}
	\acro{pea}[PEA]{pulsed-electro-acoustic method}
	\acroplural{pde}[PDEs]{partial differential equations}
	\acro{pm}[PM]{person month}
	\acroplural{pm}[PMs]{person months}
	\acro{qoi}[QoI]{quantity of interest}
	\acroplural{qoi}[QoIs]{quantities of interest}
	\acro{rdm}[RDM]{research data management}
	\acro{rms}[rms]{root mean square}
    \acro{temf}[TEMF]{Institute for Accelerator Science and Electromagnetic Fields}
	\acro{tna}[TNA]{Transient Network Analysis}
	\acro{tuda}[TUDa]{Technische~Universität~Darmstadt}
	\acro{tum}[TUM]{TU~Mün\-chen}
	\acro{uq}[UQ]{uncertainty quantification}
	\acro{wp}[WP]{work package}
	\acroplural{wp}[WPs]{work packages}
	\acro{xlpe}[XLPE]{cross-linked polyethylene}
	\acro{zno}[ZnO]{zinc oxide}
	\acro{CI}[CI]{continuous integration}
\end{acronym}

\end{document}